\documentclass[aps,pra,superscriptadress,tightenlines,floatfix,twocolumn]{revtex4-1}
\usepackage{mathrsfs}
\usepackage{graphicx}
\usepackage[english]{babel}
\usepackage{amsmath}
\usepackage{amssymb}
\usepackage{bm}% bold math

\newcommand{\Ek}{E_{\mathbf{k}}}
\newcommand{\xik}{\xi_{\mathbf{k}}}
\newcommand{\sumk}{\sum_{\mathbf{k}}}

\newcommand{\ba}{\begin{eqnarray}}
\newcommand{\ea}{\end{eqnarray}}

\begin{document}

\title{Relation connecting thermodynamics and transport of atomic unitary Fermi superfluids}

%\affiliation{Department of Physics, Southeast University, Jiulonghu Campus, Nanjing 211189, China}
%\author{Weimin Cai}
%\affiliation{Department of Physics, Southeast University, Jiulonghu Campus, Nanjing 211189, China}
%\author{Yan He}
%\affiliation{College of Physical Science and Technology, Sichuan University, Chengdu, Sichuan 610064, China}
%\author{Chih-Chun Chien}
%\affiliation{University of California Merced, School of Natural Sciences, Merced, 95343, USA.}
\author{Hao Guo$^{1}$, Weimin Cai$^1$, Yan He$^{2}$, Chih-Chun Chien$^{3\ast}$}
\affiliation{$^1$Department of Physics, Southeast University, Jiulonghu Campus, Nanjing 211189, China}
\affiliation{$^2$College of Physical Science and Technology, Sichuan University, Chengdu, Sichuan 610064, China}
\affiliation{$^3$School of Natural Sciences, University of California, Merced, CA 95343, USA}
\email{cchien5@umcerced.edu}

%%%%%%%%%%%%%%%%%%%%%%%%%%%%%%%%%%%%%%%%%%%%%%%%%%%%%%%%%%%%%%%%%%%
\begin{abstract}
The shear viscosity has been shown to be equal to the product of pressure and relaxation time in normal scale-invariant fluids, but the presence of superfluidity at low temperatures can alter the relation. By using the mean-field BCS-Leggett theory with a gauge-invariant linear response theory for unitary Fermi superfluids, we present an explicit relation between thermodynamic quantities, including the pressure and chemical potential, and transport coefficients, including the shear viscosity, superfluid density, and anomalous shear viscosity from momentum transfer via Cooper pairs. The relation is modified when pairing fluctuations associated with noncondensed Cooper pairs are considered. Within a pairing fluctuation theory consistent with the BCS-Leggett ground state, we found an approximate relation for unitary Fermi superfluids. The exact mean-field relation and the approximate one with pairing flucutaions advance our understanding of relations between equilibrium and transport quantities in superfluids, and they help determine or constrain quantities which can be otherwise difficult to measure.
\end{abstract}

\maketitle

%%%%%%%%%%%%%%%%%%%%%%%%%%%%%%%%%%%%%%%%%%%
\section{Introduction}
In physics, there are relations connecting different thermodynamic quantities in equilibrium, such as $P=(2/3)E$ for both noninteracting fermions~\cite{Walecka} and unitary Fermi gases~\cite{HoPRL04}. Here $P$ is the pressure and $E$ is the energy density, and a unitary Fermi gas consists of two-component attractive fermions on the verge of forming two-body bound states. There are also relations connecting different transport coefficients, such as the Wiedemann-Franz law showing the ratio of the thermal conductivity and the electric conductivity of normal electrons is proportional to the temperature~\cite{AshcroftBook}. There are, however, a third type of relations connecting thermodynamic quantities and transport coefficients. An example is the Einstein relation $D=\Gamma/(\partial n/\partial \mu)$ of Brownian motion~\cite{ChaikinBook}, where $D$ is the diffusion constant and is a transport coefficient, $\partial n/\partial \mu$ is the density susceptibility (which is related to the compressibility) and is a thermodynamic quantity, and $\Gamma$ is a dissipative coefficient associated with the relaxation rate of the system.

It was shown~\cite{EnssPRA12} that the shear viscosity $\eta$ is proportional to the pressure $P$ in a normal scale-invariant fluid, implying a relation $\eta=P\tau$ with $\tau$ being the relaxation time. By using a large-$N$ expansion and kinetic theory, the relation is shown to apply to normal unitary Fermi gases, where the divergent two-body $s$-wave scattering length renders the system scale invariant. The relation has been implemented in later studies of unitary Fermi gases~\cite{SchaferPRA15}. A particular feature of this relation is that $P$ is not a susceptibility, so it does not have the structure suggested by the fluctuation-dissipation theorem that leads to the Einstein relation~\cite{ChaikinBook}. The relaxation time $\tau$ naturally comes out of linear response or kinetic theories \cite{AshcroftBook,HaoPRL11,EnssPRA12}, but its measurements in cold-atoms can be challenging and past studies used the lifetime of quasi-particles as an estimation of $\tau$ \cite{HaoPRL11}. The relation $\eta=P\tau$ provides a more direct means for determining $\tau$ when $\eta$ and $P$ are measured experimentally. Since the ground state of an unitary Fermi gas is a superfluid, it is imperative to investigate how the relation is modified in the suprfluid phase.

There has been broad interest in transport and thermodynamic properties of ultracold atomic Fermi gases, which provide a clean testbed for analyzing strongly interacting systems and connect to other interacting quantum systems~\cite{KinastPRL05,BruunPRA07,SchaferPRA07,ourlongpaper,ThomasJLTP08,Nascimbene10,ThomasPRL14,ThomasScience11,PethickPRL11,HaoPRL11,ZwierleinNature11,HaoNJP11,EnssPRA12,SchaferPRA15}. The unitary Fermi gas is also closely related to the perfect quantum fluid~\cite{Turlapov08,Cao11} showing a minimal ratio between the shear viscosity and entropy. Although harmonic traps are frequently used in cold-atom experiments, recent progresses on realizing and measuring homogeneous Fermi gases~\cite{Mkherjee17,Horikoshi17} offer more direct check of many-body theories without complications from inhomogeneity.

Here we present a relation of superfluid unitary Fermi gases connecting the shear viscosity, pressure, superfluid density, chemical potential, and anomalous shear viscosity due to the order parameter. For the mean-field BCS-Leggett theory \cite{Leggett}, the relation is a consequence of the equations of state and a gauge-invariant linear response theory named the consistent fluctuations of order parameter (CFOP) theory \cite{OurPRD12,OurJLTP13}, which may also be constructed by a path integral formalism~\cite{HLYAP16}. Since strongly attractive interactions can lead to preformed pairs in unitary Fermi gases~\cite{Ourreview,OurAnnPhys}, one should distinguish condensed and noncondensed Cooper pairs. Following a pairing fluctuation theory consistent with the BCS-Leggett ground state~\cite{Ourreview,OurAnnPhys}, the noncondensed pair contributions to the thermodynamic quantities and transport coefficients in the relation are evaluated. A modified relation resembling the one in the absence of pairing fluctuations is found, and numerical calculations show that the modified relation represents a reasonable approximation.

The paper is organized as follows. Section~\ref{sec:quantities} introduces the thermodynamics quantities and transport coefficients involved in the relation of unitary Fermi superfluids. The exact relation for the BCS-Leggett theory and an approximate relation in the presence of pairing fluctuations are presented in Section~\ref{sec:relations}. The anomalous shear viscosity characterizing the momentum transfer through Cooper pairs will also be discussed there. Section~\ref{sec:conclusion} concludes our work. The theoretical framework, details, and derivations of the relations and associated quantities are summarized in the Appendix.

\section{Thermodynamic quantities and transport coefficients from BCS-Leggett theory}\label{sec:quantities}
\subsection{Mean-field theory}
The relation connecting thermodynamic and transport quantities of unitary Fermi superfluids can be derived from the mean-field BCS-Leggett theory.
The equations of states, which are usually called the gap and number equations, are given by \cite{Leggett}
$\frac{1}{g}=\sum_{\mathbf{k}}\frac{1}{2\epsilon_{\mathbf{k}}}-\frac{m}{4\pi a}=\sum_{\mathbf{k}}\frac{1-2f(E_{\mathbf{k}})}{2E_{\mathbf{k}}}$ and
$n=\sum_{\mathbf{k}}\Big[1-\frac{\xi_{\mathbf{k}}}{E_{\mathbf{k}}}+2\frac{\xi_{\mathbf{k}}}{E_{\mathbf{k}}}f(E_{\mathbf{k}})\Big]$,
where $m$ is the fermion mass, $g$ is the attractive coupling constant modeling the contact interaction between atoms, and $a$ is the two-body $s$-wave scattering length.
Here $f(x)=[\exp(x/k_B T)+1]^{-1}$ is the Fermi distribution function, $\xi_{\mathbf{k}}=\epsilon_\mathbf{k}-\mu=\frac{\hbar^2 k^2}{2m}-\mu$, and $E_{\mathbf{k}}=\sqrt{\xi^2_{\mathbf{k}}+\Delta^2}$ with $\Delta$ the order parameter or the energy gap. The chemical potential $\mu$ has to be self-consistently determined. Throughout the paper we will set $\hbar=1$ and $k_B=1$. The unitary point corresponds to $1/(k_Fa)=0$, where $k_F$ the Fermi momentum of a noninteracting Fermi gas with the same density. Via thermodynamic arguments~\cite{HoPRL04}, it has been shown that $P=\frac{2}{3}E$ for unitary Fermi gases. In the Leggett-BCS theory, the following expressions satisfy the relation $P=\frac{2}{3}E$ in the superfluid phase (with a proof in the Appendix):
\begin{eqnarray}
P&=&-\sum_{\mathbf{k}}(\xi_{\mathbf{k}}-E_{\mathbf{k}})-\frac{\Delta^2}{g}+2\sum_{\mathbf{k}}T\ln(1+e^{-\frac{E_{\mathbf{k}}}{T}}), \nonumber \\
E&=&\sum_{\mathbf{k}}(\xi_{\mathbf{k}}-E_{\mathbf{k}})+\frac{\Delta^2}{g}+2\sum_{\mathbf{k}}E_{\mathbf{k}}f(E_{\mathbf{k}})+\mu n.
\end{eqnarray}

The shear viscosity can be calculated by the Kubo formalism~\cite{Kadanoff61,BruunPRA07}:
\begin{eqnarray}\label{eta0}
\eta&=&-\lim_{\omega\rightarrow0}\lim_{q\rightarrow0}\textrm{Im}\frac{\Xi(\omega,\mathbf{q})}{\omega}=-m^2\lim_{\omega\rightarrow0}\lim_{q\rightarrow0}\frac{\omega}{q^2}\textrm{Im}\chi_{\textrm{T}}(\omega,\mathbf{q}) \nonumber \\
&=&\frac{1}{15\pi^2m^2}\int_{0}^{+\infty}dkk^6\frac{\xi^2_{\mathbf{k}}}{E^2_{\mathbf{k}}}\Big(-\frac{\partial f(E_{\mathbf{k}})}{\partial E_{\mathbf{k}}}\Big)\tau.
\end{eqnarray}
Here $\Xi(\omega,\mathbf{q})$ is the Fourier transform of  $\Xi(\bar{\tau}-\bar{\tau}',\mathbf{q})=-i\theta(\bar{\tau}-\bar{\tau}')\langle \tensor{T}^{xy}(\bar{\tau},\mathbf{q})\tensor{T}^{xy}(\bar{\tau}',-\mathbf{q})\rangle $, the stress tensor-stress tensor response function. $\bar{\tau}$ denotes the imaginary time, $\theta(x)$ is the Heaviside step function, and $\tensor{T}^{xy}$ is the $xy$-component of the energy-momentum stress tensor $\tensor{T}$. Formally,   $\tensor{T}=\frac{1}{m}\big(\nabla\psi^\dagger_{\uparrow}(\mathbf{x})\nabla\psi_{\uparrow}(\mathbf{x})+\nabla\psi^\dagger_{\downarrow}(\mathbf{x})
\nabla\psi_{\downarrow}(\mathbf{x})\big)+\tensor{I}\mathcal{L}$, where $\tensor{I}$ is the unit tensor and $\mathcal{L}$ is the Lagrangian density.
The momentum conservation law $\frac{\partial \mathbf{J}}{\partial t}+\frac{1}{m}\nabla\cdot\tensor{T}=0$ leads to the second expression of Eq.~\eqref{eta0} in terms of
the transverse current-current correlation function defined by $\chi_\textrm{T}=(\sum_{i=x}^z\chi^{ii}_{\textrm{JJ}}-\chi_{\textrm{L}})/2$. The longitudinal part is given by
$\chi_\textrm{L}=\hat{\mathbf{q}}\cdot\tensor{\chi}_{\textrm{JJ}}\cdot\hat{\mathbf{q}}$.
For BCS superfluids, the current-current response function $\tensor{\chi}_{\textrm{JJ}}$ can be evaluated from the CFOP linear response theory, and it has a tensor structure~\cite{OurPRD12,OurJLTP13}  $\tensor{\chi}_{\textrm{JJ}}=\tensor{P}+\frac{n}{m}\tensor{I}+\tensor{C}$. Here $n$ is the particle density, $\tensor{P}$ is the paramagnetic response function, and
 $\tensor{C}$ is associated with
the collective modes~\cite{KosztinPRB00,HaoPRL10} but is irrelevant to the shear viscosity~\cite{Haothesis} (see the Appendix for more details).

In Eq.~\eqref{eta0}, $\tau$ is the relaxation time measuring the broadening of the response functions. Moreover, $\eta$ is constrained by the sum rule~\cite{Kadanoff63}
$\lim_{\mathbf{q}\rightarrow\mathbf{0}}\int_{-\infty}^{+\infty}\frac{d\omega}{\pi}\Big(-\frac{\textrm{Im}\chi_{\textrm{T}}(\omega,\mathbf{q})}{\omega}\Big)=\frac{n_\textrm{n}}{m}$,
where $n_\textrm{n}$ is the normal fluid density.
Above the mean-field critical temperature $T^*$, $\Delta=0$ and Eq.~(\ref{eta0}) coincides with the expression of a normal Fermi gas~\cite{BruunPRA07}. In that case the shear viscosity reduces to
$\eta=\frac{1}{5}nv_F\tau m^\ast$, where $v_F$ is the Fermi velocity and $m^\ast$ is the effective mass (details can be found in the Appendix).

\section{Relations in unitary Fermi superfluids}\label{sec:relations}
\subsection{Relation from Leggett-BCS theory}
\begin{figure}[t]
	\centering
	\includegraphics[width=\columnwidth]{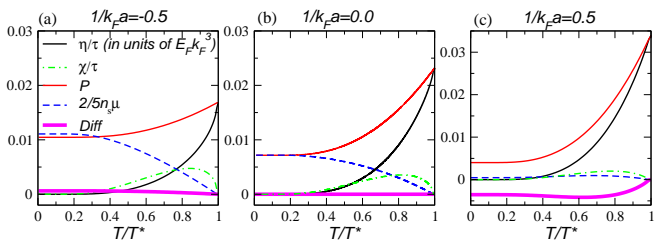}
	\caption{Plots of $\eta/\tau$ (black solid line), $\chi/\tau$ (green dot-dash line), $P$ (red solid line) and $\frac{2}{5}n_\textrm{s}\mu$ (blue dashed line) as functions of temperature (a) on the BCS side ($1/(k_F a)=-0.5$), (b) at unitary, and (c) on the BEC side ($1/(k_F a)=0.5$). The thick pink lines denote the difference,  Diff$=\eta/\tau+\chi/\tau-(P-\frac{2}{5}n_\textrm{s}\mu)$.
	}
	\label{fig:BCS}
\end{figure}
Using the BCS-Leggett theory and the CFOP linear response theory, we found a relation connecting thermodynamic quantities and transport coefficients of a mean-field unitary Fermi superfluid:
\begin{eqnarray}\label{etaP}
\eta+\chi=(P-\frac{2}{5}\mu n_s)\tau.
\end{eqnarray}
The pressure $P$ and chemical potential $\mu$ are thermodynamic quantities. On the other hand, the shear viscosity $\eta$, superfluid density $n_\textrm{s}$, and the response function $\chi$ representing the Cooper-pair contribution to momentum transfer are transport coefficients which can be inferred from linear response theory (see the Appendix for the derivations).  The two types of quantities are connected by the relaxation time $\tau$, which bears a similar structure as the Einstein relation.

A proof of the relation (\ref{etaP}) is summarized in the Appendix, and here we explain the physical meaning of the terms involved. The thermodynamic quantities such as $P$ and $\mu$ in the BCS-Legget theory has already been presented. The superfluid density can be obtained from the paramagnetic response function by \cite{Walecka,HaoNJP11}
$n_\textrm{s}=m\lim_{\omega\rightarrow0}\lim_{\mathbf{q}\rightarrow\mathbf{0}}\textrm{Re}[P^{xx}(\omega,\mathbf{q})]+n$. The response function
$\chi$ characterizes how momentum transfer can be conducted through the Cooper-pair channel in contrast to the density channel. As one will see, its derivation is almost identical to the shear viscosity except the anomalous density $\psi_{\downarrow}\psi_{\uparrow}$ replaces the normal density $\psi^{\dagger}_{\sigma}\psi_{\sigma}$ with $\sigma=\uparrow,\downarrow$. In conventional BCS theory, the anomalous Green's function also contributes to the linear response~\cite{Walecka}, and here we found another example.

Therefore, $\chi$ is identified as the anomalous shear viscosity from the Cooper-pair channel and can be obtained from the $\tensor{\Pi}$-$\tensor{\Pi}$ correlation function. Here
$\tensor{\Pi}(\mathbf{x})=\frac{1}{m}\big(\nabla\psi_{\downarrow}(\mathbf{x})\nabla\psi_{\uparrow}(\mathbf{x})+\nabla\psi^\dagger_{\uparrow}(\mathbf{x})\nabla\psi^\dagger_{\downarrow}(\mathbf{x})\big)$
is the anomalous counterpart of the stress tensor $\tensor{T}$, and the response function is
$\hat{\tensor{Q}}(\bar{\tau}-\bar{\tau}',\mathbf{q})=-i\theta(\bar{\tau}-\bar{\tau}')\langle[\tensor{\Pi}(\bar{\tau},\mathbf{q}),\tensor{\Pi}(\bar{\tau}',-\mathbf{q})]\rangle$. %where $\bar{\tau}$ is the imaginary time.
The anomalous shear viscosity $\chi$ is defined via the $xy-xy$ component of the tensor response function $\hat{\tensor{Q}}$ similar to the definition of $\eta$. Explicitly, $\chi\equiv-\lim_{\omega\rightarrow0}\lim_{q\rightarrow0}\frac{1}{\omega}\textrm{Im}[Q^{xyxy}(\omega,\mathbf{q})]$.
The final expressions in the BCS-Leggett theory are
\begin{eqnarray}\label{ns}
n_{\textrm{s}}&=&\frac{2}{3}\sum_{\mathbf{k}}\frac{\Delta^2}{E^2_{\mathbf{k}}}\frac{k^2}{m}\Big[\frac{1-2
f(E_{\mathbf{k}})}{2E_{\mathbf{k}}}+\frac{\partial f(E_{\mathbf{k}})}{\partial E_{\mathbf{k}}}\Big], \nonumber \\
\chi&=&-\frac{2}{15}\sum_{\mathbf{k}}\frac{k^4}{m^2}\frac{\Delta^2}{E^2_\mathbf{k}}\frac{\partial f(E_\mathbf{k})}{\partial E_\mathbf{k}}\tau.
\end{eqnarray}

\begin{figure}[t]
	\centering
	\includegraphics[width=\columnwidth]{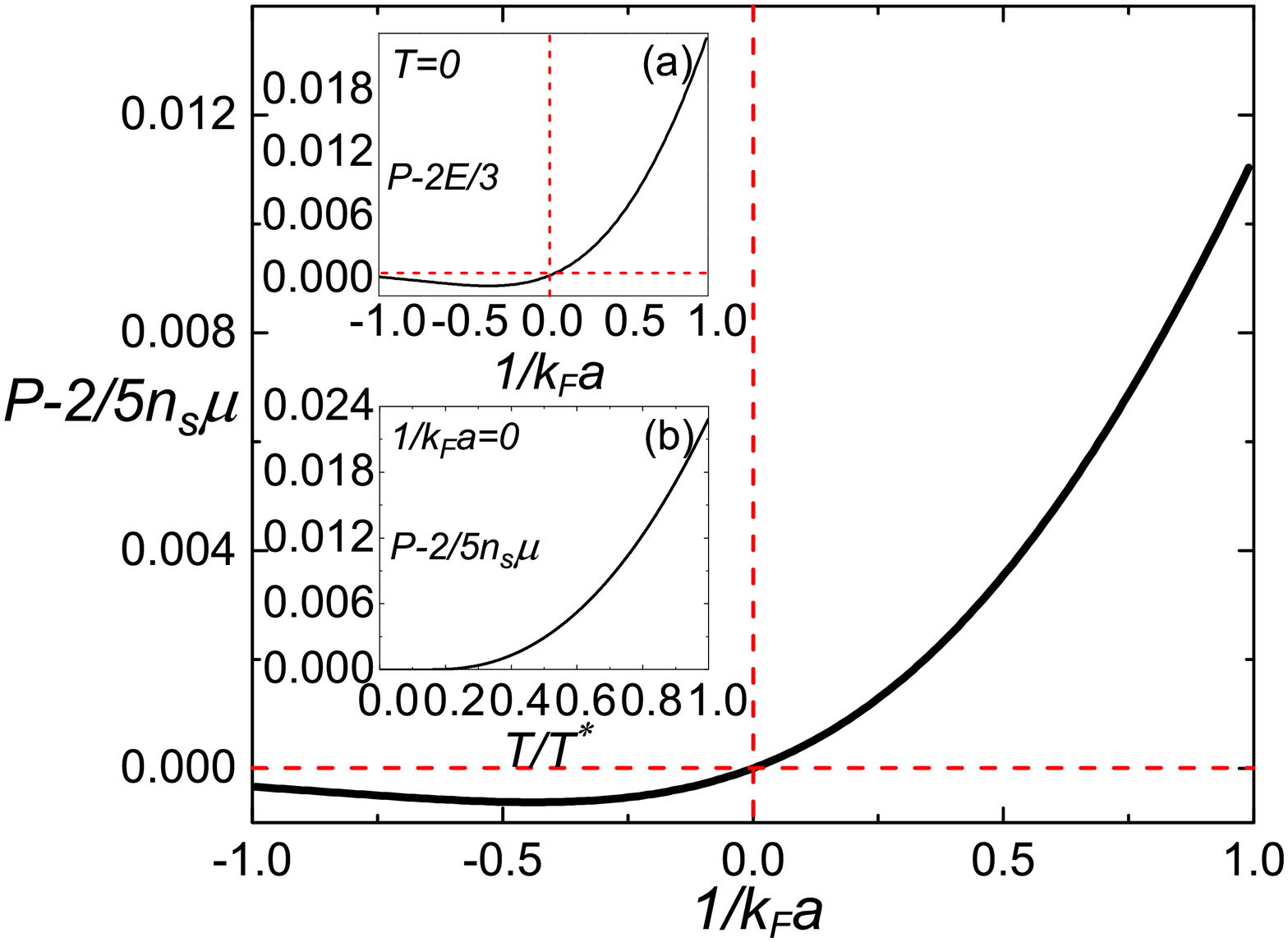}
	\caption{$P-\frac{2}{5}n_\textrm{s}\mu$ at $T=0$ as a function of $1/(k_Fa)$. Inset (a) shows  $P-\frac{2}{3}E$ vs. $1/(k_Fa)$ at zero temperature. Inset (b) shows $P-\frac{2}{5}n_s\mu$ vs. $T$ at unitarity. Here $P, E, n_\textrm{s}\mu$ are in units of $E_Fk_F^3$.
	}
	\label{diff}
\end{figure}

Above $T^*$, $n_\textrm{s}$ and $\chi$ both vanish due to the closing of the energy gap, and the relation reduces to the kinetic-theory result $\eta=P\tau$ in the normal phase. At zero temperature, $\eta$ and $\chi$ both vanish since the pure BCS ground state lacks dissipation associated with transport coefficients, and the relation reduces to
$P=\frac{2}{5}\mu n_{\textrm{s}}$, which is consistent with the experimental results of Ref.~\cite{Horikoshi17}. This zero-temperature relation remains valid when pairing fluctuations consistent with the BCS-Leggett ground state are present.

The quantities in the relation are visualized in Figure~\ref{fig:BCS}, where $\eta/\tau$, $\chi/\tau$, $P$, and $\frac{2}{5}n_\textrm{s}\mu$ as a function of temperature are shown for superfluids on the BCS side with $1/(k_Fa)=-0.5$, at unitaryity, and on the BEC side with $1/(k_Fa)=0.5$. Here $T^*/T_F=0.26, 0.50, 0.84$, respectively. The relation works well at unitarity but exhibits deviations in the other cases. However, from panel (a) this relation also works well at relatively high temperature ($0.6 T^*\lesssim T<T^*$) in the weak BCS regime. Although separate measurements of the normal and anomalous shear viscosity, $\eta$ and $\chi$, can be a challenge in experiments, they contribute to shear momentum transfer together and one may think of a composite shear viscosity $\tilde{\eta}\equiv\eta+\chi$ that accounts for the right hand side of the relation~\eqref{etaP}.

To better understand the deviation as the system moves away from the unitary point, we show in Figure~\ref{diff} the quantity $P-\frac{2}{5}n_\textrm{s}\mu$ at zero temperature as a function the interaction strength $1/(k_Fa)$ in the regime $-1\le1/(k_Fa)\le1$. At unitarity, the quantity vanishes as indicated by the relation \eqref{etaP}. We also show, in the insets, $P-\frac{2}{3}E$ as a function of $1/(k_Fa)$ at zero temperature and $P-\frac{2}{5}n_s\mu$ as a function of $T$ at $1/(k_Fa)=0$. The qualitative behavior of $P-\frac{2}{5}n_\textrm{s}\mu$ is similar to $P-\frac{2}{3}E$ at zero temperature, which implies that the deviation is associate with the scale-invariance which guarantees certain thermodynamic relations. Moreover, $P-\frac{2}{5}n_\textrm{s}\mu$ is close to $0$ already at low $T$, and this may ease its verification in experiments. Figures~\ref{fig:BCS} and \ref{diff} also suggest that even though the relation \eqref{etaP} is not exact on the BCS side, the deviation is relatively minor and one may continue using it there as an approximation.

\subsection{Pairing fluctuation effects}
\subsubsection{Beyond mean-field theory}
Due to the strongly attractive interactions in unitary Fermi gases, pairing fluctuation effects discerning the gap function and the order parameter should be considered ~\cite{Ourreview,OurAnnPhys,OurNSR10,OurJLTP13,ourlongpaper}. Moreover, the onset temperature of pairing $T^*$ and that of condensation (or superfluid transition temperature) $T_c$ are different, and from $t$-matrix calculations $T^*/T_F$ can be more than twice as large as $T_c/T_F$ \cite{OurAnnPhys,OurNSR10}. Here we follow a $t$-matrix theory consistent with the BCS-Leggett ground state and derive the thermodynamics first. The Green's function has the generic form $G(P)=[G_0^{-1}(P)-\Sigma(P)]^{-1}$, where $G_0$ is the noninteracting Green's function. The self energy comes from ladder-diagram corrections to the propagator and is given by $\Sigma(P)=\sum_{Q}t(Q)G_0(P-Q)$, where the $t$-matrix $t(Q)$ may be separated into the condensed ($Q=0$) and noncondensed ($Q\neq 0$) pair contributions. Thus,  $t(Q)\approx t_\textrm{sc} + t_\textrm{pg}$ with $t_{\textrm{sc}}(Q)=-(\Delta_{\textrm{sc}}^2/T)\delta(Q)$ and $t_\textrm{pg}(Q)=[g^{-1}+X(Q)]^{-1}$ is constructed from $X(Q)=\sum_K G_0(Q-K)G(K)$ which consists of one bare and one full Green's functions.
Here $K=(i\omega_n,\mathbf{k})$ and $Q=(i\Omega_l,\mathbf{q})$ are fermionic and bosonic  four-momenta, respectively. The total gap can also be decomposed into the order parameter (from the condensed pairs) $\Delta_\textrm{sc}$ and the pseudogap (from the noncondensed pairs) $\Delta_\textrm{pg}$ as $\Delta^{2}=\Delta_{\textrm{sc}}^{2}+\Delta_{\textrm{pg}}^{2}$, where the pseudogap is approximated by $\Delta_\textrm{pg}^2=-\sum_{Q}t_{\textrm{pg}}(Q)$, and the superfluid transition temperature is determined by where $\Delta_\textrm{sc}$ vanishes. The modified number and gap equations can be derived from $n=\sum_P G(P)$ and $\frac{1}{g}=\sum_PG(P)G_0(-P)$. The pairs have a finite lifetime, which contributes to the relaxation time $\tau$ in linear response.

The density of grand potential, $\Omega=-P$, can be derived from the same framework \cite{ourlongpaper}, and it leads to the aforementioned equations of states after minimization. There are also contributions from a fermionic part (including the condensed pairs and quasi-particles) and from a bosonic part (from the noncondensed pairs).  Explicitly,
$P=-\Omega=-(\Omega_\textrm{f}+\Omega_\textrm{b})=P_\textrm{f}+P_\textrm{b}$,
where $\Omega_\textrm{f}=\Delta^2\chi_0+\sum_{\mathbf{k}}\big[(\xik-\Ek)-2T\ln(1+e^{-\frac{\Ek}{T}})\big]$ and $\Omega_\textrm{b}=T\sum_{\mathbf{q}}\ln(1-e^{-\frac{\Omega_\mathbf{q}}{T}})$.
Here $\chi_0=\frac{1}{g}-Z\mu_\textrm{pair}$, $\Omega_\mathbf{q}\approx\frac{q^2}{2M^*}-\mu_\textrm{pair}$, $Z=\frac{\partial X}{\partial \Omega}|_{\Omega=0,\mathbf{q}=0}$, $\mu_\textrm{pair}$ is the pair chemical potential, and $M^*=12(\frac{\partial^2\chi(Q)}{\partial k^2}|_{Q=0})^{-1}$ is the effective pair mass. The total energy can be decomposed in a similar fashion into
$E_\textrm{f}=\sum_{\mathbf{k}}(\xi_{\mathbf{k}}-E_{\mathbf{k}})+\Delta^2\chi_0+2\sum_{\mathbf{k}}E_{\mathbf{k}}f(E_{\mathbf{k}})+\mu n$ and
$E_\textrm{b}=\sum_\mathbf{q}(\Omega_\mathbf{q}+\mu_\textrm{pair})b(\Omega_\mathbf{q})$,
where $b(x)=[\exp(x/T)-1]^{-1}$ is the Bose distribution function. Similar to the mean-field theory results, for unitary Fermi gases one can show that $E_\textrm{f}=\frac{3}{2}P_\textrm{f}$ and $E_\textrm{b}=\frac{3}{2}P_\textrm{b}$, so the relation $E=\frac{3}{2}P$ still holds when pairing fluctuations are included.

The transport coefficients such as the shear viscosity can be evaluated from a modified CFOP theory with additional diagrams included  to ensure the Ward identities~\cite{HaoPRL10,HaoPRL11}, which in turn guarantees charge conservation.  Under a weak dissipation assumption~\cite{KosztinPRB00,HaoPRL10,Ourreview} the paramagnetic response function $\tensor{P}(\omega,\mathbf{q})$ can be evaluated and the shear viscosity can be derived accordingly (with the derivations summarized in the Appendix). An important observation is that since the pressure $P$ in Eq.~(\ref{etaP}) receives corrections from the noncondensed pairs (acting like composite bosons), it is natural to modify the shear viscosity to acquire similar (fermionic and bosonic) corrections, $\eta=\eta_\textrm{f}+\eta_\textrm{b}$. The former comes from a calculation similar to the BCS-Leggett theory, and the latter may be approximated by the shear viscosity of a free Bose gas:
\begin{eqnarray}\label{etapg}
\eta_\textrm{f}&=&\int_{0}^{+\infty}dk\frac{k^6}{15\pi^2m^2}\frac{E^2_{\mathbf{k}}-\Delta^2_\textrm{pg}}{E^2_{\mathbf{k}}}\frac{\xi^2_{\mathbf{k}}}{E^2_{\mathbf{k}}}\Big(-\frac{\partial f(E_{\mathbf{k}})}{\partial E_{\mathbf{k}}}\Big)\tau, \nonumber \\
\eta_\textrm{b}&=&-\frac{1}{30\pi^2M^{\ast2}}\int_0^\infty dkk^6\frac{\partial b(\Omega_\mathbf{k})}{\partial \Omega_\mathbf{k}}\tau.
\end{eqnarray}
Since the noncondensed pairs are in local equilibrium with the fermions, it may be reasonable to assume that the relaxation time for composite bosons is the same as that for fermions. Moreover, we found that $\eta_\textrm{b}=P_\textrm{b}\tau$, so in this approximation the noncondensed pairs satisfy the simple relation for normal scale-invariant systems.

\begin{figure}[t]
	\centering
	\includegraphics[width=\columnwidth]{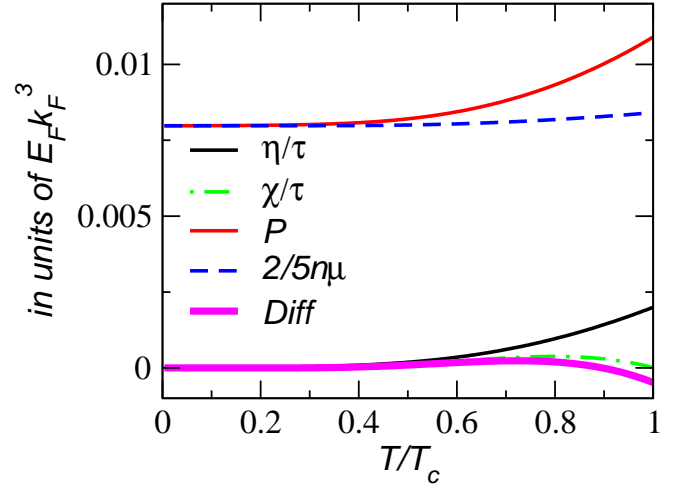}
	\caption{$\eta/\tau$ (black line), $\chi/\tau$ (green dot-dash line), pressure $P$ (red line), $\frac{2}{5}n\mu$ (blue dashed line) and the difference (thick pink line), and  Diff=$(\eta+\chi)/\tau-(P-(2/5)n\mu)$ of a unitary Fermi superfluid as a function of temperature with pairing fluctuation effects included.
	}
	\label{fig:PG}
\end{figure}

The anomalous shear viscosity $\chi$ is due to momentum transfer through the correlations among Cooper pairs, and the phase rigidity of superfluidity is essential. Therefore, according to Eq.~\eqref{ns} its expression should be modified as
$\chi=-\frac{2}{15}\sum_{\mathbf{k}}\frac{k^4}{m^2}\frac{\Delta^2_\textrm{sc}}{E^2_\mathbf{k}}\frac{\partial f(E_\mathbf{k})}{\partial E_\mathbf{k}}\tau$. This approximation also avoids possible double-counting of the shear viscosity from noncondensed pairs (already included in $\eta_b$).
The terms on the right-hand side (RHS) of the relation (\ref{etaP}) should also receive bosonic corrections from the noncondensed pairs. In the pairing fluctuation theory consistent with the BCS-Leggett ground state~\cite{OurAnnPhys}, there is no noncondensed pairs at zero temperature, so the zero-temperature relation $P=(2/5)\mu n_s$ should remain the same.
To include the noncondensed-pair contributions at finite temperatures, we replace the superfluid density $n_\textrm{s}$ on the RHS of Eq.~\eqref{etaP} by the total number density $n$ since the two components (the superfluid and the noncondensed pairs) both contribute to thermodynamics, but $n_s=n$ at $T=0$ and this is consistent with the BCS-Legget ground state. Finally, we collect all the terms and modify the relation (\ref{etaP}) of a unitary Fermi superfluid as
 \begin{eqnarray}\label{etaPpg}
\tilde{\eta}\equiv\eta+\chi\approx(P-\frac{2}{5}\mu n)\tau.
\end{eqnarray}

To verify the validity of this approximate relation, we perform numerical calculations to check each quantity. Figure~\ref{fig:PG} shows $\eta/\tau$, $\chi/\tau$, $P$, and $\frac{2}{5}n\mu$ as a function of temperature below the superfluid transition temperature $T_c$ for a unitary Fermi superfluid, with the convention of the labels the same as Figure~\ref{fig:BCS}. Here $T_c/T_F\approx 0.26$, and it is known that the $t$-matrix approximation overestimates the transition temperature~\cite{OurAnnPhys}, which should be around $0.16T_F$ in experiments~\cite{Nascimbene10}. The difference $(\eta+\chi)/\tau+\frac{2}{5}n\mu-P$ is also shown. For $T/T_c\le 0.5$, there is no observable difference, and the maximal ratio between the difference and the pressure $P$ in the regime $0.5\le T/T_c \le 1$ is less than $4.4\%$. Hence, the relation (\ref{etaPpg}) works reasonably for unitary Fermi superfluids and may serve as a condition for constraining the relevant thermodynamic and transport  quantities.

\section{Conclusion}\label{sec:conclusion}
To summarize, we have presented a relation connecting thermodynamic quantities and transport coefficients of unitary Fermi superfluids. The relation involves the shear viscosity, pressure, superfluid density, particle density, anomalous shear viscosity characterizing momentum transfer via the Cooper pairs, chemical potential, and relaxation time. The relation is exact for the mean-field BCS Leggett theory, and an approximate relation is found when pairing fluctuations from noncondensed Cooper pairs are considered. The relations illustrate interesting properties of scale-invariant quantum fluids.

Our results of unitary Fermi superfluids also contrast different contributions from the condensate, fermionic quasiparticles, and noncondensed pairs and their interplays. The relations helps constrain elusive physical quantities such as the relaxation time if other quantities (the pressure, shear viscosity, etc.) can be measured experimentally. Moreover, the theoretical framework presented here may inspire searches for similar relations between thermodynamic quantities and transport coefficients in other many-body systems.

\textit{Acknowledgment:} We thank Qijin Chen for useful discussions. H. G. thanks the support from the National Natural Science Foundation of China (Grant No. 11674051).

\appendix
\section{Gauge-invariant linear response theory}
Here we summarize the gauge-invariant linear response theory implemented in this work, from which the transport coefficients can be deduced. Details of the theory can be found in Refs.~\cite{OurPRD12,OurJLTP13}. We start with the Leggett-BCS theory and then generalize it by including pairing fluctuation effects.
The density-current linear response is constructed by perturbing the system with an external (effective) electromagnetic (EM) field, whose potential $A^\mu$ is associated with an U(1) gauge symmetry. A self-consistent theory must maintain the gauge symmetry. The current  $J^\mu=(\rho,\mathbf{J})$, where $\rho$ is the mass density and $\mathbf{J}$ is the mass current, receives a perturbation
$\delta J^\mu(Q)=K^{\mu\nu}(Q)A_\nu(Q)$,
where $K^{\mu\nu}$ is the response function and $Q$ is the external 4-momentum.

By introducing the bare EM interaction vertex  $\gamma^\mu(K+Q,K)=(1,\frac{\mathbf{k}+\frac{\mathbf{q}}{2}}{m})$, the response function can be evaluated by
$K^{\mu\nu}(Q)=2\sum_K\big(\Gamma^{\mu}(K+Q,K)G(K+Q)\gamma^{\nu}(K,K+Q)G(K)\big)+\frac{n}{m}h^{\mu\nu}$,
where $\Gamma^{\mu}$ is the full EM interaction vertex, $h^{\mu\nu}=-\eta^{\mu\nu}(1-\eta^{\nu0})$ with $\eta^{\mu\nu}=\textrm{diag}(1,-1,-1-1)$, and $G(K)$ is the full fermion Green's function with $K$ being the fermionic 4-momentum. If the perturbed current is also conserved, we have $\partial_\mu \delta J^\mu=q_{\mu}\delta J^\mu(Q)=0$, where $q_\mu\equiv Q$. This can be achieved by requiring the EM vertex to satisfy the Ward identity $q_{\mu}\Gamma^{\mu}(K+Q,K)=G^{-1}(K+Q)-G^{-1}(K)$ and maintain the U(1) gauge symmetry, so the full vertex can be gauge invariant. Nambu proposed an integral-equation approach~\cite{Nambu60}, but its exact expression remains unknown.

Here we construct a gauge-invariant vertex by applying the consistent fluctuations of order parameter (CFOP) theory~\cite{Arseev,OurJLTP13} and obtain
\begin{eqnarray}\label{V1}\Gamma^{\mu}(K+Q,K)&=&\gamma^{\mu}(K+Q,K)+\textrm{Coll}_\textrm{sc}^\mu(K+Q,K) \nonumber \\
& &+\textrm{MT}_\textrm{sc}^\mu(K+Q,K).
 \end{eqnarray}
The second term is associated with the contributions from the Nambu-Goldstone (or collective) mode, but the expression is irrelevant to the relation discussed here (see Ref.~\cite{Haothesis} for the details and a summary below). The third term is the Maki-Thompson (MT) diagram associated with the condensate, $\label{MTSC}\textrm{MT}_\textrm{sc}^\mu(K+Q,K)=-\Delta^2 G_0(-K)\gamma^{\mu}(-K,-K-Q)G_0(-K-Q)$ with $G_0(K)$ denoting the bare (noninteracting) fermionic Green's function. After plugging those expressions in $K^{\mu\nu}$, the spatial part is given by $\tensor{\chi}_{\textrm{JJ}}\equiv \tensor{K}=\tensor{P}+\frac{\tensor{n}}{m}+\tensor{C}$. The last two terms can be shown to be irrelevant to the evaluation of shear viscosity, and the expression of $\tensor{P}$ will be given in a moment. In the mean-field theory, $\Delta$ is identical to the order parameter.

After considering pairing fluctuation effects, one should also modify the vertex (\ref{V1}) accordingly to maintain the Ward identity. A gauge invariant solution is
\begin{eqnarray}\label{V2}
\Gamma^{\mu}(P+Q,P)&=&\gamma^{\mu}(P+Q,P)+\textrm{Coll}^{\mu}_{\textrm{sc}}(P+Q,P) \nonumber \\
& &+\textrm{MT}^{\mu}_{\textrm{sc}}(P+Q,P)+\textrm{MT}^{\mu}_{\textrm{pg}}(P+Q,P) \nonumber \\
& &+\textrm{AL}^{\mu}_{1}(P+Q,P)+\textrm{AL}^{\mu}_{2}(P+Q,P).
\end{eqnarray}
Here $\textrm{MT}^{\mu}_{\textrm{pg}}$ denotes the MT diagram for non-condensed pairs and AL$_{1,2}$ denote two Aslamazov-Larkin (AL) diagrams. The AL diagrams are present because non-condensed pairs carry momentum, and they are absent in the mean-field theory.
The expressions of the vertices are given by
\begin{align}&\textrm{MT}^{\mu}_{\textrm{pg}}(P+Q,P)=\sum_Kt_{\textrm{pg}}(K)G_{0}(K-P)\notag\\
&\times\gamma^{\mu}(K-P,K-P-Q)G_{0}(K-P-Q),\notag\\
&\textrm{AL}^{\mu}_{1}(P+Q,P)=-\sum_{K,L}t_{\textrm{pg}}(K)t_{\textrm{pg}}(K+Q)\notag\\
&\times G_{0}(K-P)G(K-L)G_{0}(L+Q)\gamma^{\mu}(L+Q,L)G_{0}(L),\notag\\
&\textrm{AL}^{\mu}_{2}(P+Q,P)=-\sum_{K,L}t_{\textrm{pg}}(K)t_{\textrm{pg}}(K+Q)\notag\\
&\times G_{0}(K-P)G_{0}(K-L)G(L+Q)\Gamma^{\mu}(L+Q,L)G(L).
\end{align}
Moreover, they satisfy the relation
\begin{eqnarray}
&q_{\mu}\big[\frac{1}{2}\textrm{AL}^{\mu}_{1}(P+Q,P)+\frac{1}{2}\textrm{AL}^{\mu}_{2}(P+Q,P)\notag\\
&+\textrm{MT}^{\mu}_{\textrm{pg}}(P+Q,P)\big]=0,
\end{eqnarray}
which guarantees gauge-invariance.

After collecting the modifications, the spatial part of the response function has a similar structure
$\tensor{\chi}_{\textrm{JJ}}=\tensor{P}+\frac{n}{m}\tensor{I}+\tensor{C}$, and again one can show that only
$\tensor{P}$ is relevant to the shear viscosity.

\section{Details of the results from BCS-Leggett theory}
\subsection{Viscosity}
%\begin{widetext}
The paramagnetic response function from the CFOP theory can be derived from the formalism shown in Ref.~\cite{KulikJLTP81}, and its explicit expression is
\begin{align}\label{P}
\tensor{P}(\omega,\mathbf{q})&=\sum_{\mathbf{k}}\frac{\mathbf{k}\mathbf{k}}{m^2}\Big\{\big(1-\frac{\xi^+_{\mathbf{k}}\xi^-_{\mathbf{k}}+\Delta^2}{E^+_{\mathbf{k}}E^-_{\mathbf{k}}}\big)
\notag\\&\times\frac{E^+_{\mathbf{k}}+E^-_{\mathbf{k}}}{\omega^2-(E^+_{\mathbf{k}}+E^-_{\mathbf{k}})^2}[1-f(E^+_{\mathbf{k}})-f(E^-_{\mathbf{k}})] \notag \\
& -\big(1+\frac{\xi^+_{\mathbf{k}}\xi^-_{\mathbf{k}}+\Delta^2}{E^+_{\mathbf{k}}E^-_{\mathbf{k}}}\big)\notag\\&\times
\frac{E^+_{\mathbf{k}}-E^-_{\mathbf{k}}}{\omega^2-(E^+_{\mathbf{k}}-E^-_{\mathbf{k}})^2}[f(E^+_{\mathbf{k}})-f(E^-_{\mathbf{k}})]\Big\},
\end{align}
%\end{widetext}
where $\xi^{\pm}_{\mathbf{k}}=\xi_{\mathbf{k}\pm\frac{\mathbf{q}}{2}}$ (the convention also holds for $E^{\pm}_{\mathbf{k}}$).
%Only $\tensor{P}$ has the non-zero transverse part in the small $\omega$ and $q$ limit.
To evaluate the angular part of the integral, we choose the direction of $\mathbf{q}$ as the direction of the $z$-axis and let the angle between $\mathbf{k}$ and $\mathbf{q}$ be $\theta$. In the $\omega\rightarrow0$, $q\rightarrow0$ limit, the shear viscosity shown in Eq. (2) in the main text  is given by
\begin{eqnarray}\label{eta10}
& &\eta=-m^2\lim_{\omega\rightarrow0}\lim_{q\rightarrow0}\frac{\pi\omega}{2q^2}\sum_{\mathbf{k}}\frac{k^2\textrm{sin}^2\theta}{m^2}\nonumber\\
&\times&\Big[\frac{E^+_{\mathbf{k}}E^-_{\mathbf{k}}-\xi^+_{\mathbf{k}}\xi^-_{\mathbf{k}}-\Delta^2}{2E^+_{\mathbf{k}}E^-_{\mathbf{k}}}\big(1-f(E^+_{\mathbf{k}})-f(E^-_{\mathbf{k}})\big)\nonumber\\
&\times&\big(\delta(\omega+E^+_{\mathbf{k}}+E^-_{\mathbf{k}})-\delta(\omega-E^+_{\mathbf{k}}-E^-_{\mathbf{k}})\big)\nonumber\\
&-&\frac{E^+_{\mathbf{k}}E^-_{\mathbf{k}}+\xi^+_{\mathbf{k}}\xi^-_{\mathbf{k}}+\Delta^2}{2E^+_{\mathbf{k}}E^-_{\mathbf{k}}}\big(f(E^+_{\mathbf{k}})-f(E^-_{\mathbf{k}})\big)\nonumber\\&\times&\big(\delta(\omega+E^+_{\mathbf{k}}-E^-_{\mathbf{k}})-\delta(\omega-E^+_{\mathbf{k}}+E^-_{\mathbf{k}})\big)\Big].
\end{eqnarray}The $\delta$-functions in the third line vanish since $E^+_{\mathbf{k}}+E^-_{\mathbf{k}}>2\Delta\neq\omega\rightarrow0$. By using $\lim_{q\rightarrow0}(E^+_{\mathbf{k}}-E^-_{\mathbf{k}})=\mathbf{q}\cdot\nabla E_{\mathbf{k}}=\frac{\mathbf{k}\cdot\mathbf{q}}{m}\frac{\xi_{\mathbf{k}}}{E_{\mathbf{k}}}$, we obtain
\begin{eqnarray} \label{eta4}
\eta&=&-\lim_{\omega\rightarrow0}\lim_{q\rightarrow0}\frac{1}{30\pi m^2}\int_0^{\infty}dkk^6\frac{\xi^2_{\mathbf{k}}}{E^2_{\mathbf{k}}}\frac{\partial f(E_{\mathbf{k}})}{\partial E_{\mathbf{k}}}\nonumber\\&\times&\big(\delta(\omega+\mathbf{q}\cdot\nabla E_{\mathbf{k}})+\delta(\omega-\mathbf{q}\cdot\nabla E_{\mathbf{k}})\big).
\end{eqnarray}
Here we include dissipation effects by replacing the $\delta$-functions with Lorentzian functions:
\begin{eqnarray} \label{RG1}
\delta(x)=\lim_{\Gamma\rightarrow0}\frac{1}{\pi}\frac{\Gamma}{x^2+\Gamma^2}\textrm{ with } \tau=\frac{1}{\Gamma},
\end{eqnarray}
then we get
\begin{eqnarray}
\eta=\frac{1}{15\pi^2m^2}\int_{0}^{+\infty}dkk^6\frac{\xi^2_{\mathbf{k}}}{E^2_{\mathbf{k}}}\Big(-\frac{\partial f(E_{\mathbf{k}})}{\partial E_{\mathbf{k}}}\Big)\tau.
\end{eqnarray}
When $T>T^*$, $\Delta=0$, then in the BCS limit
\begin{eqnarray}
\eta
&=&\frac{2\tau}{15m^2}\int\frac{d^3\mathbf{k}}{(2\pi)^3}\big(-\frac{\partial f(\xi_{\mathbf{k}})}{\partial \xi_{\mathbf{k}}}\big)\nonumber\\
&\simeq&\frac{8\tau_{\eta}}{15}N(0)\int_{-\infty}^{+\infty}d\xi_{\mathbf{k}}(\xi_{\mathbf{k}}+\mu)^2\big(-\frac{\partial f(\xi_{\mathbf{k}})}{\partial \xi_{\mathbf{k}}}\big).
\end{eqnarray}
where $N(0)=\frac{mk_F}{2\pi^2}$ is the density of states near the Fermi surface. Since the derivative of Fermi function is an even function,
\begin{eqnarray}
\eta&\simeq&\frac{16}{15}N(0)\tau\Big(\int_0^{+\infty}d\xi_{\mathbf{k}}\xi_{\mathbf{k}}^2\big(-\frac{\partial f(\xi_{\mathbf{k}})}{\partial \xi_{\mathbf{k}}}\big)\nonumber\\
& &+\int_0^{+\infty}d\xi_{\mathbf{k}}\mu^2\big(-\frac{\partial f(\xi_{\mathbf{k}})}{\partial \xi_{\mathbf{k}}}\big)\Big)\nonumber\\
&=&\frac{8\tau}{15}N(0)\mu^2(1+\frac{T^2\pi^2}{3\mu^2})\nonumber\\
&=&\frac{1}{5}nv^2_F\tau m^\ast,
\end{eqnarray}
where $n=\frac{k^3_F}{3\pi^2}$ is the particle density, $v_F=\frac{k_F}{m}$, $m^\ast=m(1+\frac{T^2\pi^2}{3\mu^2})$, and we have applied $\mu\approx\frac{k^2_F}{2m}$ in the BCS limit.

In the main text, we claim the terms $\frac{n}{m}\tensor{I}$ and $\tensor{C}$ shown in $\tensor{\chi}_{\mathbf{J}\mathbf{J}}$ give no contribution to the shear viscosity. The former has no imaginary part, so it does not enter the expression of $\eta$. The details of the latter can be found in Refs.~\cite{Haothesis,OurJLTP13}, and here we give a brief explanation. Assuming the direction of $\mathbf{q}$ is along the $z$-axis, in the small-$\omega$, $q$ limit it can be shown that only the component $C^{zz}$ does not vanish. Next, one can show that $\lim_{\omega\rightarrow0}\lim_{q\rightarrow 0}\sum_{i=x}^z\chi^{ii}_{\textrm{JJ}}=\lim_{\omega\rightarrow0}\lim_{q\rightarrow 0}C_\textrm{L}$. According to the definition of the transverse component of a tensor, $\lim_{\omega\rightarrow0}\lim_{q\rightarrow 0}C_\textrm{T}$ vanishes. A direct calculation then shows that $\tensor{C}$ does not contribute to the shear viscosity. This applies to both mean-field and pairing-fluctuation theories.
In retrospect, the shear viscosity comes from the response function of the energy-momentum stress tensor, which is determined by the space-time structure and the Hamiltonian. On the other hand, the
collective modes, or Goldstone modes, are from symmetry breaking of the phase of order parameter and do not couple directly to the stress tensor calculations.

\subsection{Relation between $E$ and $P$ for unitary Fermi superfluids}\label{appb}
Here we present a proof of $E=\frac{3}{2}P$ explicitly for unitary Fermi superfluids at the mean-field level. Since $\frac{1}{k_Fa}=0$, the coupling constant should satisfy   $\frac{1}{g}=\sum_{\mathbf{k}}\frac{1}{2\epsilon_{\mathbf{k}}}$. Then,
\begin{eqnarray}\label{Ef}
& &E-\frac{3}{2}P \\
&=&\sumk(\xik-\Ek+\frac{\Delta^2}{2\epsilon_{\mathbf{k}}})+\sumk2\Ek f(\Ek)+\mu n\nonumber\\
&+&\frac{3}{2}\sumk(\xik-\Ek+\frac{\Delta^2}{2\epsilon_{\mathbf{k}}})-3\sumk T\ln(1+e^{-\Ek/T}).\nonumber
\end{eqnarray}
Applying integration by parts,
\ba\label{t1}
\sumk T\ln(1+e^{-\Ek/T})=\frac{1}{3}\sumk\frac{k^2}{m}\frac{\xik}{\Ek}f(\Ek).
\ea
With Eq.~(\ref{Ef}), one obtains
$\left(\mu\frac{2\xik}{\Ek}-\frac{k^2}{m}\frac{\xik}{\Ek}+2\Ek\right)f(\Ek)
=\frac{2\Delta^2}{\Ek}f(\Ek)$.
The third term on the right hand side of Eq.~(\ref{Ef}) can also be rewritten as
$
\frac{3}{2}\sumk(\xik-\Ek+\frac{\Delta^2}{2\epsilon_{\mathbf{k}}})=-\sumk \frac{k^2}{2m}\big(1-\frac{\xik}{\Ek}-\frac{\Delta^2}{2\epsilon^2_{\mathbf{k}}}\big)
$.
Collect all terms of Eq.~(\ref{Ef}) which do not contain $f(\Ek)$, we have
\ba
& &\sumk(\xik-\Ek+\frac{\Delta^2}{2\epsilon_{\mathbf{k}}})+\mu\sumk\left(1-\frac{\xik}{\Ek}\right)\nonumber\\
&-&\sumk \frac{k^2}{2m}\left(1-\frac{\xik}{\Ek}-\frac{\Delta^2}{2k^4}\right)\nonumber\\
&=&\Delta^2\left(\frac{1}{\epsilon_{\mathbf{k}}}-\frac{1}{\Ek}\right).
\ea
Now putting everything together we have
\ba
E-\frac{3}{2}P=\Delta^2\sumk\left(\frac{1}{\epsilon_{\mathbf{k}}}-\frac{1}{\Ek}+\frac{2f(\Ek)}{\Ek}\right)=0,
\ea
where the gap equation $\frac{1}{g}=\sum_{\mathbf{k}}\frac{1}{2\epsilon_{\mathbf{k}}}-\frac{m}{4\pi a}=\sum_{\mathbf{k}}\frac{1-2f(E_{\mathbf{k}})}{2E_{\mathbf{k}}}$ has been applied. In the presence of pairing fluctuations, a similar derivation can be performed and the relation $P=(2/3)E$ still holds~\cite{ourlongpaper}, and more details will be shown in Appendix~\ref{appe}.

\subsection{Anomalous shear viscosity}
The shear viscosity, in classical picture, quantifies the momentum transfer perpendicular to the direction of an applied force. In linear response theory of quantum fluids, it corresponds to the response function due to perturbations of the energy-momentum stress tensor. By replacing the normal density/current by the anomalous density/current from Cooper pairs, the same linear repsonse theory leads to a response function characterizing the momentum transfer through the Cooper-pair channel. In momentum space we introduce the anomalous current operator $\tensor{\Pi}$, which has a tensor form:
$\tensor{\Pi}(\mathbf{q})=\sum_\mathbf{k}\frac{\mathbf{k}(\mathbf{k}+\mathbf{q})}{m}(\psi_{-\mathbf{k}\downarrow}\psi_{\mathbf{k}+\mathbf{q}\uparrow}+\psi^\dagger_{\mathbf{k}\uparrow}\psi^\dagger_{-\mathbf{k}-\mathbf{q}\downarrow})$,
By implementing the Nambu-Gorkov spinors
\begin{equation}\label{Ns1}
\Psi_{\mathbf{k}}=\left[\begin{array}{c}
\psi_{\mathbf{k}\uparrow} \\
\psi^{\dagger}_{-\mathbf{k}\downarrow}\end{array}\right], \qquad
\Psi^{\dagger}_{\mathbf{k}}=[\psi^{\dagger}_{\mathbf{k}\uparrow},\psi_{-\mathbf{k}\downarrow}],
\end{equation}
the anomalous tensor current can be expressed by
\begin{align}
\tensor{\Pi}(\mathbf{q})=\sum_\mathbf{k}\Psi^\dagger_{\mathbf{k}}\tensor{\gamma}(\mathbf{k},\mathbf{k}+\mathbf{q})\Psi_{\mathbf{k}+\mathbf{q}},
\end{align}
where $\tensor{\gamma}(\mathbf{k},\mathbf{k}+\mathbf{q})=\frac{\mathbf{k}(\mathbf{k}+\mathbf{q})}{m}\sigma_1$, with the Pauli matrix $\sigma_1=\left(\begin{array}{cc} 0 & 1\\ 1 & 0\end{array}\right)$, is a generalized dyadic interaction vertex. We assume $\tensor{\Pi}$ couples to an external tensor field $\tensor{\Phi}$, so the perturbed Hamiltonian becomes
\begin{equation}
H'=\sum_\mathbf{q}\tensor{\Pi}(\mathbf{q}):\tensor{\Phi}(\mathbf{q})=\sum_{\mathbf{k}\mathbf{q}}\Psi^\dagger_{\mathbf{k}}\tensor{\gamma}(\mathbf{k},\mathbf{k}+\mathbf{q}):\tensor{\Phi}(\mathbf{q})\Psi_{\mathbf{k}+\mathbf{q}},
\end{equation}
where the operation $:$ denotes a contraction between tensors, or a dyadic product in the situation here. The anomalous tensor current perturbed by the external field can be evaluated by linear response theory. We introduce the imaginary time $\bar{\tau}=it$,
and the perturbed anomalous tensor current is defined by
\begin{equation}
\delta\tensor{\Pi}(\bar{\tau},\mathbf{q})=\sum_\mathbf{k}\langle\Psi^\dagger_{\mathbf{k}}(\bar{\tau})\tensor{\gamma}(\mathbf{k},\mathbf{k}+\mathbf{q})\Psi_{\mathbf{k}+\mathbf{q}}(\bar{\tau})\rangle,
\end{equation}
and in linear response theory it becomes
$\delta\tensor{\Pi}(\bar{\tau},\mathbf{q})=\hat{\tensor{Q}}(\bar{\tau}-\bar{\tau}',\mathbf{q}):\tensor{\Phi}(\bar{\tau}',\mathbf{q})$,
where the response function $\hat{\tensor{Q}}(\bar{\tau}-\bar{\tau}',\mathbf{q})$ is a fourth-order tensor with the expression
\begin{equation}
\hat{\tensor{Q}}(\bar{\tau}-\bar{\tau}',\mathbf{q})=-i\theta(\bar{\tau}-\bar{\tau}')\langle[\tensor{\Pi}(\bar{\tau},\mathbf{q}),\tensor{\Pi}(\bar{\tau}',-\mathbf{q})]\rangle.
\end{equation}
After a Fourier transform, the response function becomes
%\begin{widetext}
\begin{align}
\hat{\tensor{Q}}(\omega,\mathbf{q})&=\sum_{\mathbf{k}}\frac{\mathbf{k}^-\mathbf{k}^+\mathbf{k}^+\mathbf{k}^-}{m^2}\Big\{\big(1+\frac{\xi^+_{\mathbf{k}}\xi^-_{\mathbf{k}}-\Delta^2}{E^+_{\mathbf{k}}E^-_{\mathbf{k}}}\big)\notag\\
&\times\frac{E^+_{\mathbf{k}}+E^-_{\mathbf{k}}}{\omega^2-(E^+_{\mathbf{k}}+E^-_{\mathbf{k}})^2}\big[1-f(E^+_{\mathbf{k}})-f(E^-_{\mathbf{k}})\big] \notag \\
&-\big(1-\frac{\xi^+_{\mathbf{k}}\xi^-_{\mathbf{k}}-\Delta^2}{E^+_{\mathbf{k}}E^-_{\mathbf{k}}}\big)\notag\\
&\times\frac{E^+_{\mathbf{k}}-E^-_{\mathbf{k}}}{\omega^2-(E^+_{\mathbf{k}}-E^-_{\mathbf{k}})^2}\big[f(E^+_{\mathbf{k}})-f(E^-_{\mathbf{k}})\big]\Big\}.
\end{align}
%\end{widetext}
The anomalous shear viscosity $\chi$ is obtained in the same way via the response function:
\begin{align}
\chi&=-\lim_{\omega\rightarrow0}\lim_{q\rightarrow0}\frac{1}{\omega}\textrm{Im}Q^{xyxy}(\omega,\mathbf{q})\notag\\
&=-\lim_{\omega\rightarrow0}\lim_{q\rightarrow0}\frac{\pi}{\omega}\sum_{\mathbf{k}}\frac{\mathbf{k}^{-x}\mathbf{k}^{+y}\mathbf{k}^{-y}\mathbf{k}^{+x}}{m^2}\notag\\
&\times\Big[\frac{E^+_{\mathbf{k}}E^-_{\mathbf{k}}+\xi^+_{\mathbf{k}}\xi^-_{\mathbf{k}}-\Delta^2}{2E^+_{\mathbf{k}}E^-_{\mathbf{k}}}
\big(1-f(E^+_{\mathbf{k}})-f(E^-_{\mathbf{k}})\big)\notag\\
&\times\big(\delta(\omega+E^+_{\mathbf{k}}+E^-_{\mathbf{k}})-\delta(\omega-E^+_{\mathbf{k}}-E^-_{\mathbf{k}})\big)\notag\\
&-\frac{E^+_{\mathbf{k}}E^-_{\mathbf{k}}-\xi^+_{\mathbf{k}}\xi^-_{\mathbf{k}}+\Delta^2}{2E^+_{\mathbf{k}}E^-_{\mathbf{k}}}\big(f(E^+_{\mathbf{k}})-f(E^-_{\mathbf{k}})\big)
\notag\\
&\times\big(\delta(\omega+E^+_{\mathbf{k}}-E^-_{\mathbf{k}})-\delta(\omega-E^+_{\mathbf{k}}+E^-_{\mathbf{k}})\big)\Big]\notag\\
&=-\frac{2}{15}\sum_{\mathbf{k}}\frac{k^4}{m^2}\frac{\Delta^2}{E^2_\mathbf{k}}\frac{\partial f(E_\mathbf{k})}{\partial E_\mathbf{k}}\tau.
\end{align}
Here $Q^{xyxy}$ is the $xyxy$ component of the tensor $\hat{\tensor{Q}}$, and Eq.~(\ref{RG1}) has been used to regularize the $\delta$-function. Since the arguments of the $\delta$-functions here are the same as those appearing in the expression of $\eta$ in Eq.~(\ref{eta10})), we can assume that the relaxation time for $\chi$ is the same as that for $\eta$.

\subsection{Proof of the main relation}\label{appc}
First we derive the superfluid density defined by \cite{Walecka,HaoNJP11}
\begin{align}\label{nsP}
n_\textrm{s}=m\lim_{\omega\rightarrow0}\lim_{\mathbf{q}\rightarrow\mathbf{0}}\textrm{Re}[P^{xx}(\omega,\mathbf{q})]+n.
\end{align}
Using integration by parts, the total particle density is given by
\begin{align}\label{nsc}
n&=\sum_{\mathbf{k}}\Big(1-\frac{\xi_{\mathbf{k}}}{E_{\mathbf{k}}}\big[1-2f(E_{\mathbf{k}})\big]\Big)\notag\\
&=\int^{+\infty}_0\frac{dk}{2\pi^2}k^2\Big(1-\frac{\xi_{\mathbf{k}}}{E_{\mathbf{k}}}\big[1-2f(E_{\mathbf{k}})\big]\Big)\nonumber \\
&=\frac{2}{m}\sum_{\mathbf{k}}\frac{k^2}{3E^2_{\mathbf{k}}}\Big(\frac{1-2f(E_{\mathbf{k}})}{2E_{\mathbf{k}}}\Delta^2-\xi^2_{\mathbf{k}}\frac{\partial f(E_{\mathbf{k}})}{\partial E_{\mathbf{k}}}\Big).
\end{align}
The first term on the RHS of Eq.~(\ref{nsP}) is evaluated as
\begin{align}\label{nsP1}
m\lim_{\omega\rightarrow0}\lim_{\mathbf{q}\rightarrow\mathbf{0}}\textrm{Re}[P^{xx}(\omega,\mathbf{q})]
%=\sum_{\mathbf{k}}\frac{k^xk^x}{m}\frac{\partial f(E_{\mathbf{k}})}{\partial E_{\mathbf{k}}}\notag\\
=\frac{2}{3}\sum_{\mathbf{k}}\frac{k^2}{m}\frac{\partial f(E_{\mathbf{k}})}{\partial E_{\mathbf{k}}}.
\end{align}
Therefore,
\begin{align}\label{nsc3}
n_\textrm{s}=\frac{2\Delta^2}{3m}\sum_{\mathbf{k}}\frac{k^2}{E^2_{\mathbf{k}}}\Big(\frac{1-2f(E_{\mathbf{k}})}{2E_{\mathbf{k}}}+\frac{\partial f(E_{\mathbf{k}})}{\partial E_{\mathbf{k}}}\Big).
\end{align}

Next we briefly derive Eq.~(4) in the main text. By integrating Eq.~(3) in the main text by parts, we get
\begin{eqnarray}\label{eta1}
\eta&=&\frac{1}{3\pi^2}\int_{0}^{+\infty}dk\frac{k^4}{m}\frac{\xi_{\mathbf{k}}}{E_{\mathbf{k}}}f(E_{\mathbf{k}})\tau\nonumber\\&+&
\frac{1}{15\pi^2m}\int_{0}^{+\infty}dk\frac{k^6}{m}\frac{\Delta^2}{E^3_{\mathbf{k}}}f(E_{\mathbf{k}})\tau,
\end{eqnarray}
Comparing this equation with Eq.~(\ref{t1}) and applying the expression of $P$ (see Eq.~(1) in the main text), we get
\begin{eqnarray}\label{eta2}
\eta&=&P\tau+\sum_{\mathbf{k}}(\xi_{\mathbf{k}}-E_{\mathbf{k}}+\frac{\Delta^2}{2\epsilon_{\mathbf{k}}})\tau\nonumber\\
&+&\frac{1}{15\pi^2m}\int_{0}^{+\infty}dk\frac{k^6}{m}\frac{\Delta^2}{E^3_{\mathbf{k}}}f(E_{\mathbf{k}})\tau.
\end{eqnarray}
Applying $E=\frac{3}{2}P$ and Eq.(\ref{t1}), we have
\begin{eqnarray}
& &\sum_{\mathbf{k}}(\xi_{\mathbf{k}}-E_{\mathbf{k}}+\frac{\Delta^2}{2\epsilon_{\mathbf{k}}})\\
&=&-\frac{2}{5}\Big(-3\sum_{\mathbf{k}}T\ln(1+e^{-\frac{E_{\mathbf{k}}}{T}})+2\sum_{\mathbf{k}}E_{\mathbf{k}}f(E_{\mathbf{k}})+\mu n\Big)\nonumber\\
&=&-\frac{2}{5}\sum_{\mathbf{k}}\Big[2\frac{\Delta^2}{E_{\mathbf{k}}}f(E_{\mathbf{k}})-2\mu\frac{\xi_{\mathbf{k}}}{E_{\mathbf{k}}}f(E_{\mathbf{k}})+\mu n\Big]\nonumber\\
&=&-\frac{2}{5}\sum_{\mathbf{k}}\Big[2\frac{\Delta^2}{E_{\mathbf{k}}}f(E_{\mathbf{k}})+\mu\big(1-
\frac{\xi_{\mathbf{k}}}{E_{\mathbf{k}}}\big)\Big]\nonumber\\
&=&-\frac{1}{15\pi^2m}\int_{0}^{+\infty}dk\frac{k^6}{m}\frac{\Delta^2}{E^3_{\mathbf{k}}}f(E_{\mathbf{k}})\nonumber\\
&-&\frac{1}{15\pi^2}\int_{0}^{+\infty}dk\frac{\Delta^2}{E^2_{\mathbf{k}}}\frac{k^4}{m}\Big[\mu\frac{1-2
f(E_{\mathbf{k}})}{E_{\mathbf{k}}}-2\xi_{\mathbf{k}}\frac{\partial f(E_{\mathbf{k}})}{\partial E_{\mathbf{k}}}\Big].\nonumber
\end{eqnarray}
Substitute this result to Eq.~(\ref{eta2}), we finally get
\begin{eqnarray}\label{etaP2}
\eta&=&P\tau\nonumber\\&-&\frac{2}{15\pi^2}\int_{0}^{+\infty}dk\frac{\Delta^2}{E^2_{\mathbf{k}}}\frac{k^4}{m}\Big[\mu\frac{1-2
f(E_{\mathbf{k}})}{2E_{\mathbf{k}}}-\xi_{\mathbf{k}}\frac{\partial f(E_{\mathbf{k}})}{\partial E_{\mathbf{k}}}\Big]\tau\nonumber\\
&=&P\tau-\frac{2}{5}\mu n_{\textrm{s}}\tau+\frac{2}{15}\sum_{\mathbf{k}}\frac{\Delta^2}{E^2_{\mathbf{k}}}\frac{k^4}{m^2}\frac{\partial f(E_{\mathbf{k}})}{\partial E_{\mathbf{k}}}\tau.
\end{eqnarray}
The last term is just $-\chi$, so
\begin{equation}
\eta+\chi=(P-\frac{2}{5}\mu n_s)\tau.
\end{equation}

\section{Results from pairing fluctuation theory}
\subsection{Noncondensed pair contribution to the shear viscosity}\label{appe}
The paramagnetic response kernel in the presence of pairing fluctuations is given by \cite{HaoPRL10}
\begin{eqnarray}\label{P1}
& &\overset\leftrightarrow{P}(\omega,\mathbf{q})=\sum_{\mathbf{k}}\frac{\mathbf{k}\mathbf{k}}{m^2}\Big\{\big(1-\frac{\xi^+_{\mathbf{k}}\xi^-_{\mathbf{k}}+\Delta^2_\textrm{sc}-\Delta^2_\textrm{pg}}{E^+_{\mathbf{k}}E^-_{\mathbf{k}}}\big)\nonumber\\
&\times&\frac{E^+_{\mathbf{k}}+E^-_{\mathbf{k}}}{\omega^2-(E^+_{\mathbf{k}}+E^-_{\mathbf{k}})^2}[1-f(E^+_{\mathbf{k}})-f(E^-_{\mathbf{k}})] \nonumber \\
& &-\big(1+\frac{\xi^+_{\mathbf{k}}\xi^-_{\mathbf{k}}+\Delta^2_\textrm{sc}-\Delta^2_\textrm{pg}}{E^+_{\mathbf{k}}E^-_{\mathbf{k}}}\big)\nonumber \\&\times&\frac{E^+_{\mathbf{k}}-E^-_{\mathbf{k}}}{\omega^2-(E^+_{\mathbf{k}}-E^-_{\mathbf{k}})^2}[f(E^+_{\mathbf{k}})-f(E^-_{\mathbf{k}})]\Big\}.
\end{eqnarray}
The fermionic part of the shear viscosity can be derived in a way similar to the mean-field theory. For the bosonic (non-condensed pair) part, we consider a free boson gas with the Hamiltonian
\begin{align}\label{Hb}
H_\textrm{b}=\sum_\mathbf{q}\Omega_\mathbf{q}b^\dagger_\mathbf{q}b_\mathbf{q},
\end{align}
where $\Omega_\mathbf{q}\approx\frac{q^2}{2M^*}-\mu_\textrm{pair}$ is the bosonic energy dispersion with $M^*$ the effective boson mass. The bosonic Green's function is
\begin{align}
G_\textrm{b}(i\Omega_l,\mathbf{q}) =\frac{1}{i\Omega_l-\Omega_\mathbf{q}}.
\end{align}
The bosonic current operator is
\begin{align}
\mathbf{J}_{\textrm{b}}(\bar{\tau},\mathbf{q})=-\frac{1}{M^*}\sum_\mathbf{k}(\mathbf{k}+\frac{\mathbf{q}}{2})b^\dagger_\mathbf{k}(\bar{\tau})b_{\mathbf{k}+\mathbf{q}}(\bar{\tau}).
\end{align}
This defines a $\mathbf{J}-\mathbf{J}$ linear response, and the response function is given by
\begin{equation}
\tensor{Q}_\textrm{b}^{\mathbf{J}\mathbf{J}}(\bar{\tau}-\bar{\tau}',\mathbf{q})=-i\theta(\bar{\tau}-\bar{\tau}')\langle[\mathbf{J}_{\textrm{b}}(\bar{\tau},\mathbf{q}),\mathbf{J}_{\textrm{b}}(\bar{\tau}',-\mathbf{q})]\rangle.
\end{equation}
After a Fourier transform,
\begin{align}
\tensor{Q}_\textrm{b}^{\mathbf{J}\mathbf{J}}(Q)&=-T\sum_{i\omega_m}\sum_\mathbf{k}\frac{\mathbf{k}+\frac{\mathbf{q}}{2}}{M^*}\frac{\mathbf{k}+\frac{\mathbf{q}}{2}}{M^*}G_\textrm{b}(K+Q)G_\textrm{b}(K)\notag\\
&=-\sum_\mathbf{k}\frac{\mathbf{k}\mathbf{k}}{M^{\ast2}}\frac{b(\Omega^+_\mathbf{k})-b(\Omega^-_\mathbf{k})}{i\Omega_l-\Omega^+_\mathbf{k}+\Omega^-_\mathbf{k}},
\end{align}
where $\Omega^\pm_\mathbf{k}=\Omega_{\mathbf{k}\pm\frac{\mathbf{q}}{2}}$, $Q=(i\Omega_l,\mathbf{q})$ and $K=(i\omega_m,\mathbf{k})$ are both bosonic four-momenta. Following the analytic continuation $i\Omega_l\rightarrow \omega+i0^+$, the shear viscosity is given by
\begin{align}
\eta_\textrm{b}&=-M^{\ast2}\lim_{\omega\rightarrow0}\lim_{q\rightarrow0}\textrm{Im}\frac{\omega}{q^2}Q^{\mathbf{J}\mathbf{J}}_{\textrm{b}\textrm{T}}(\omega,\mathbf{q})\notag\\
&=-\frac{1}{30\pi^2M^{\ast2}}\int_0^\infty dkk^6\frac{\partial b(\Omega_\mathbf{k})}{\partial \Omega_\mathbf{k}}\tau.
\end{align}
Here we have assumed that the boson relaxation time is the same as the fermion relaxation time.
Finally, we found the following relation for the noncondensed pairs:
\begin{align}
P_\textrm{b}=\frac{2}{3}E_\textrm{b}=\frac{\eta_\textrm{b}}{\tau}.
\end{align}
This can be proved in two steps:
\begin{align}
P_\textrm{b}&=-\frac{T}{2\pi^2}\int^\infty_0dqq^2\ln(1-e^{-\frac{\Omega_\mathbf{q}}{T}})\notag\\
%&=-\frac{T}{6\pi^2}\int^\infty_0 dq^3\ln(1-e^{-\frac{\Omega_\mathbf{q}}{T}})\notag\\
&=-\frac{T}{6\pi^2}q^3\ln(1-e^{-\frac{\Omega_\mathbf{q}}{T}})\Big|^\infty_0\notag\\&+\frac{T}{6\pi^2}\int^\infty_0dqq^3\frac{-e^{-\frac{\Omega_\mathbf{q}}{T}}}{1-e^{-\frac{\Omega_\mathbf{q}}{T}}}\big(-\frac{1}{T}\frac{q}{M^*}\big)\notag\\
&=\frac{2}{3}\sum_\mathbf{q}\frac{q^2}{2M^*}\frac{1}{e^{\frac{\Omega_\mathbf{q}}{T}}-1}\notag\\
&=\frac{2}{3}E_\textrm{b}.
\end{align}
Since $E_\textrm{f}$ and $P_\textrm{f}$ of the fermionic part resemble the mean-field expressions  given in Appendix~\ref{appb}, it can be shown that $P_\textrm{f}=\frac{2}{3}E_\textrm{f}$. Hence, the relation $P=\frac{2}{3}E$ still holds when pairing fluctuation effects are considered.
In the second step, integration by parts leads to
\begin{align}
P_\textrm{b}&=\frac{1}{30\pi^2M^*}\int^\infty_0 dq^5\frac{1}{e^{\frac{\Omega_\mathbf{q}}{T}}-1}\notag\\
&=\frac{1}{30\pi^2M^*}q^5\frac{1}{e^{\frac{\Omega_\mathbf{q}}{T}}-1}\Big|^\infty_0\notag\\
&-\frac{1}{30\pi^2M^*}\int^\infty_0 dqq^5(-1)\frac{e^{\frac{\Omega_\mathbf{q}}{T}}}{(e^{\frac{\Omega_\mathbf{q}}{T}}-1)^2}\frac{1}{T}\frac{q}{M^*}\notag\\
&=-\frac{1}{30\pi^2M^{\ast2}}\int_0^\infty dqq^6\frac{\partial b(\Omega_\mathbf{q})}{\partial \Omega_\mathbf{q}}\notag\\
&=\frac{\eta_\textrm{b}}{\tau}.
\end{align}

\bibliographystyle{apsrev4-1}
%\bibliography{Review,Review1}
% \bibliography{All,addon}

\end{document}